# VUV Processing of Nitrile Ice: Direct Comparison of Branching in Ice and TPD Spectra

Travis J. Hager, Bailey M. Moore, Quentin D. Borengasser, Kyle T. Renshaw, Rachel Johnson, Anudha C. Kanaherarachchi, Bernadette M. Broderick*

*Department of Chemistry, University of Missouri, Columbia MO 65211*

Correspondence: broderickbm@missouri.edu

# Abstract

The interplay between radiation chemistry and sublimation dynamics of condensed organic compounds on cold grains is fundamental to describe observed gas-phase and ice-phase molecular abundances in the interstellar medium (ISM). Infrared measurements are generally used to identify molecules synthesized in irradiated ices in laboratory experiments, while mass spectrometric techniques have been used to monitor the products following temperature programmed desorption. The IR measurements are often used quantitatively to monitor chemical transformation of ices during the course of irradiation, but the gas phase methods applied with TPD generally do not permit quantitative branching determination. Here we combine reflection-absorption infrared spectroscopy (RAIRS) of ices with broadband rotational spectroscopy of the sublimed products, to study the branching of photoproducts produced by the VUV (120 – 160 nm) irradiation of condensed $CH_3CN$ (methyl cyanide) and $CH_3CH_2CN$ (ethyl cyanide) ices. This permits direct comparison between the ice-phase and gas-phase branching following temperature programmed desorption (TPD). This comparison is analogous to astronomical observations of ices in protostellar disks such as by the James Webb Space Telescope employed in conjunction with ALMA observations in the corresponding warm-up regions of the same objects. In the condensed $CH_3CN$ VUV processed ices we quantified the HCN, $CH_3NC$ (methyl isocyanide), $CH_2CCNH$ (ketenimine), $CH_3NH_2$ (methylamine), and $CH_4$ (methane) abundances. The $CH_3CH_2CN$ ices also readily produced the corresponding isocyanide and HCN in addition to a significant yield of $CH_2CHCN$ (vinyl cyanide). The ethyl cyanide ice produced $CH_3CHNH$ (methyl ketenimine) rather than $CH_3NH_2$ and no $CH_4$ formation was observed. In the gas phase we detected the isocyanides, HCN and vinyl cyanide. The ice and gas phase relative abundances could all be brought into agreement if the unknown IR band strength of the isocyanides C-N stretch is assumed to be ~3 times larger than that of the cyanides.

Table of Contents Figure

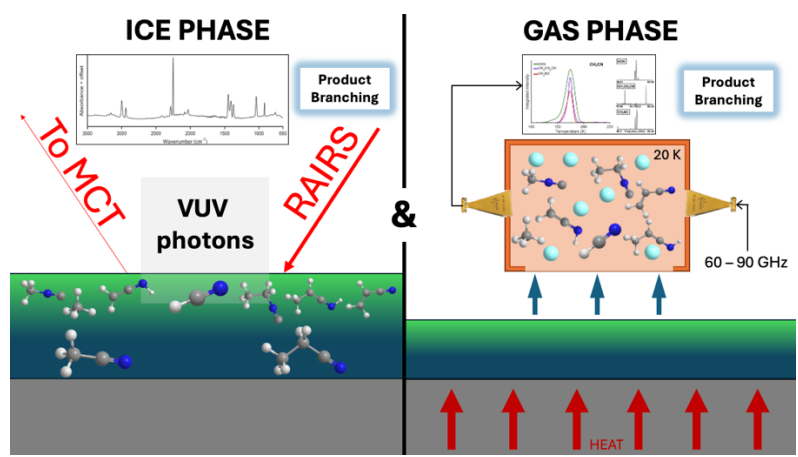

# Introduction

The freeze-out and desorption cycles of molecules on interstellar grains within star-forming regions is a fundamental process in the evolution of complex organic molecules (COMs) in the interstellar medium (ISM). In these regions, volatiles condense onto cold submicron-sized dust grains (T = ~ 10 K)[1] where the bulk of the chemistry that forms these larger molecules occurs. VUV photons, via the Prasad-Tarafdar mechanism, and cosmic rays penetrate deep within dense molecular clouds, bombarding icy-covered grains to produce reactive radicals, H-atoms, and ionic species to serve as reactants in the ultimate formation of larger COMs.[2, 3] As these grains move to warmer regions, the variety of products may sublime into the gas-phase and may be subsequently detected via Earth and space-based telescopes.[4-8] The composition of the ices formed on icy grains depends on several parameters including the unique deposition and sublimation temperatures of molecular species. This phenomenon accounts for the observation of molecular snowlines, where the desorption of a particular volatile species correlates with radial distance from the disk center, with grains experiencing warmer temperatures closer to the forming star. Temperature also impacts species reactivity within the ice, affecting both thermal and non-thermal processes.[9-14] Thus, the relationship between desorption and condensed-phase reactivity is reflected in the ratio of gas-phase to solid-phase products.

Astronomical investigations have in turn focused on the relationship between observed molecular abundances in different phases.[15-21]. Most recently, the James Webb Observations of Young protoStars (JOYS+) survey compared column density ratios of various oxygen-bearing COMs (O-COMs) relative to methanol ($CH_3OH$) in both the gas and ice phases in the IRAS 2A and B1-c protostellar sources.[17] Gas-phase column densities were obtained from mm-wave spectra observed by the Atacama Large Millimeter/submillimeter Array (ALMA), while condensed-phase column densities were derived from spectra obtained from the James Webb Space Telescope (JWST) Mid-InfraRed Instrument-Medium Resolution Spectroscopy (MIRI-MRS). It was found that the ratios of $CH_3OCHO$ and $CH_3OCH_3$ to $CH_3OH$ are consistent between the ice and gas phases. However, the ratios of $CH_3CHO$ and $CH_3CH_2OH$ differ by 1-2 orders of magnitude with significantly higher abundances observed in the ice.

Although laboratory experiments often combine IR observation of ices with gas phase measurements following TPD, in general only the IR measurements are used for branching determination as mass spectrometric detection is difficult to apply in a quantitative way, and gas phase and ice measurements have not been correlated. An exception is the work of Bergantini et al. who used tunable VUV photoionization to compare ethanol and dimethyl ether yields in irradiated methane/water ices.[22] In that case distinct ionization thresholds and known photoionization cross sections permitted a comparison of the branching for formation of the two isomers. The use of rotational spectroscopy here provides a more general, quantitative "fingerprint" detection of sublimed molecules directly analogous to that employed in astronomical observations. We here determine the relative product yields among several key products in the ice and the TPD spectra for the first time.

Surveys of nitrogen-bearing COMs(N-COMs), though less abundant than O-COMs, have also been conducted. JWST offers unprecedented sensitivity for the direct detection of N-COMs in the condensed phase, enabling more precise determinations of molecular abundances. As a result, both methyl cyanide ($CH_3CN$) and ethyl cyanide ($CH_3CH_2CN$) have been tentatively identified in ice using JWST.[18] The column density of these species were compared between the gas and ice phases across several protostellar systems. It was found that the condensed-phase ratio of 0.1 for $CH_3CH_2CN:CH_3CN$ was in good agreement with gas-phase ratios in these regions.[18] However, significant variations were observed in the $OCN^-$, $CH_3CH_2CN$, and $CH_3CN$ ratios relative to methanol. The relative ratios of vinyl cyanide ($CH_2CHCN$) and $CH_3CH_2CN$ with $CH_3CN$ have also been reported for several other interstellar regions. In Sagittarius B2N, these ratios were found to be 0.4 and 0.9, respectively, while in Orion KL the ratios were found to be a comparable 0.5 and 1.1. These values differ from the ratios found in the quiescent giant molecular cloud, G+0.693, where this ratio was found to be 0.8 and 0.4.[23, 24] The column density ratios of gas-phase O- and N-containing species have been shown to be consistent across several protostellar systems and comets.[15, 25] These findings highlight the uncertainties associated with both the chemical and physical processes involved in the production and sublimation of COMs, and in turn provide an opportunity for laboratory studies to assist in rationalizing these astronomical observations.

Prior to the observations of JWST, Saturn's largest moon, Titan, prompted extensive research into nitrile radiation chemistry. Titan's nitrogen-rich atmosphere, with a minor hydrocarbon component, contains numerous N-COMs in both gas and aerosol phases.[26-30] $CH_3CN$ was the first molecule detected in Titan's atmosphere using mm-wave wavelengths in ALMA, and it is the second most abundant nitrile in the ISM. In protoplanetary disks and comets, its abundance is ~ 0.01% relative to water in gas phase measurements, but an upper limit of 1.2% was determined in ices.[15, 17] $CH_3CH_2CN$ has also been detected in Titan's atmosphere by ALMA, but a distinct distribution and associated chemistry are inferred.[28] According to Loison et al., these nitriles are strongly correlated in their formation pathways on Titan, with both relying on the presence of the $CH_2CN$ fragment.[29] However, the model greatly overestimates $CH_3CH_2CN$ abundances, particularly in the lower atmosphere and the predictions deviate from previous models by several orders of magnitude. Cordiner et al. conclude their paper with a call for a better understanding of $CH_3CH_2CN$ formation pathways on Titan.[28]

This topic, and nitrogen's role in amino acid formation, has inspired a vast body of laboratory research focused on condensed-phase nitrile chemistry.[31-40] These studies have contributed to the development of astrochemical models to further understand the interplay of conditions that lead to the formation of COMs.[10, 29] Laboratory experiments have primarily explored ice-phase chemistry induced by energetic processing with infrared spectroscopy[41, 42] (FTIR, RAIRS or transmission) and gas-phase analysis via mass spectrometry[43-45] (laser-desorption, and photoionization). While these sensitive techniques enable the detection of COMs formed in low abundances, they often provide limited molecule-specific information due to the need for deconvolution of overlapping spectral features. This limitation hinders the differentiation of species with similar functional groups including isomers and conformers, and makes complex mixture analysis difficult.

Recent advances in mm-wave spectroscopy have enabled the laboratory study of molecules, including products formed from energetic processing, sublimed from ices with high structural specificity.[46, 47] One such instrument, Chirped Pulse ICE (CPICE), combines broadband mm-wave rotational spectroscopy with a buffer gas cell (BGC), enhancing sensitivity to warm molecules that have thermally desorbed from an ice. The approach has demonstrated the ability to quantify isomer and conformer distributions in complex systems, probe the vibrational temperature of polyatomic molecules over the course of sublimation, and observe the desorption of products formed via electron and VUV irradiation of ices.[48-51] In the present study, CPICE, with new *in-situ* FTIR capabilities, is used to quantify the branching ratios of the photoproducts generated by broadband VUV irradiation of two N-COMs, methyl cyanide (MeCN) and ethyl cyanide (EtCN). Temperature-programmed desorption (TPD) experiments are conducted to examine the distinct desorption profiles of various daughter species produced by VUV processing of these two different parent cyanides. The condensed phase abundances of VUV photoproducts measured with IR spectroscopy can be directly compared to the gas-phase abundances measured with broadband rotational spectroscopy following thermal desorption. This report presents the first combined measurement of MeCN and EtCN VUV photoproduct branching ratios *in-situ* and following thermal desorption from the ice.

**Methods**

CPICE is an ultrahigh vacuum (UHV, $1 \times 10^{-10}$ Torr) stainless steel chamber within which gas phase molecules are deposited onto a silver substrate at low temperatures to form an ice, as shown in Figure 1. The neat or mixed ices may then be irradiated using either an electron gun or a VUV lamp. The products of this irradiation are detected following sublimation over the course of TPD in a 20 K buffer gas cell (BGC) utilizing broadband rotational spectroscopy in the 60-90 GHz regime. A detailed description of this instrument has been previously provided.[50, 52]

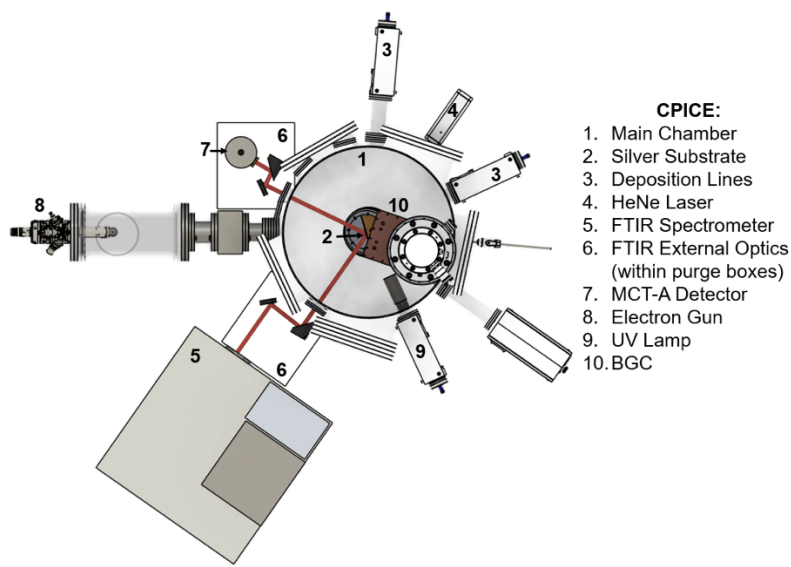

**Figure 1.** Top-down schematic of CPICE with labels in the RAIRS measurement configuration.

CPICE:
1. Main Chamber
2. Silver Substrate
3. Deposition Lines
4. HeNe Laser
5. FTIR Spectrometer
6. FTIR External Optics (within purge boxes)
7. MCT-A Detector
8. Electron Gun
9. UV Lamp
10. BGC

**Ice Formation** MeCN (Sigma-Aldrich >99% purity), and EtCN (Sigma-Aldrich >99% purity) are purified through a series of three freeze, pump, thaw (FPT) cycles before deposition. Following the third FPT, an ice is generated by slowly depositing room-temperature vapor of the parent species onto a 1 cm x 1 cm silver substrate cooled via an attached 4 K coldhead (Sumitomo

RDK415D2-F70L) mounted to the bottom of the CPICE apparatus. In these experiments, the substrate was set to 20 K throughout deposition and irradiation to allow for increased mobility of reactants and in turn more observed variety of products. Ice thickness is quantified through laser interferometry with a 632.8 nm HeNe laser during deposition, allowing for highly accurate measurement of ice thickness at the nanometer scale as discussed previously in greater detail.[50]

**VUV processing and RAIRS FTIR.** Following a 1.6 um thick vapor deposition, the ices are irradiated with a broadband VUV $D_2$ lamp (Hamamatsu L10706) which emits photons with a spectral wavelength range of 120 - 165 nm (10.3 – 7.5 eV). At 160 nm, the lamp emits with 80% irradiance, and 20% irradiance at 120 nm corresponding to a photon flux of approximately $\varphi = 10^{15}$ photons cm$^{-2}$ s$^{-1}$. The ices are irradiated for a total of 3 hours, leading to a final fluence of 2.2 x$10^{18}$ photons cm$^{-2}$. Assuming the photon attenuation within the ice follows the Beer-Lambert law, the VUV absorption cross section is used to determine the depth of photon penetration and the total dose of radiation within the penetrating region. Due to the lack of literature values of this cross-section of both condensed phase MeCN and EtCN, the averaged absorption cross-section of gas-phase MeCN in the 120-165 nm region that has been scaled to the known relationship between the gas and solid-phase absorption cross section of methanol is used. This is the same technique used by Bulak et. al. to calculate their penetration depth for their lamp in an acetonitrile ice.[53] A VUV absorption cross section of 7.5 x $10^{-18}$ cm$^2$ is used. The penetration depth is calculated to be 0.5 μm in our nitrile ices. The number of parent molecules within this irradiated volume of ice given the irradiated area and penetration depth is calculated to obtain the dose in eV per molecule. In this work, the density of amorphous MeCN (0.78 g cm$^{-3}$) and EtCN (0.70 g cm$^{-3}$) at 15 K as reported by Hudson is employed to determine the final dose after irradiation.[54] For MeCN, this was determined to be 130 eV molecule$^{-1}$, while for EtCN we compute a dosage of 195 eV molecule$^{-1}$ after 3 hours. To compare these irradiation conditions to conditions found in the interstellar medium, the secondary photon flux is estimated to be on the order of $10^4$ photons cm$^{-2}$ s$^{-1}$ [3] within a dense molecular cloud. This leads to a total fluence of 3 – 9 x $10^{18}$ photons cm$^{-2}$ over the 10 – 30 Myr lifetime of dense molecular clouds.[55] A fluence between 0 – 11 x $10^{18}$ photons cm$^{-2}$ were investigated in this work, consistent with interstellar conditions.

Recently, a Fourier Transform Infrared (FTIR) Spectrometer (Thermo Nicolette iS20) was incorporated into the CPICE apparatus for in-situ ice measurements and in turn permit the detection of synthesized products over the course of irradiation to understand when products are formed and/or destroyed. In the experiments described in this report, the spectrometer takes an IR scan every 0, 0.25, 0.5, 1, 1.5, 2, and 3 hour elapsed irradiation time points. Each IR scan is an average of 64 scans sweeping from 650 – 4000 cm$^{-1}$ at a rate of 1.5 scans per second with a 4 cm$^{-1}$ resolution. The substrate must be rotated to a pre-established angle facing the IR source for proper reflection into the liquid-$N_2$ cooled mercury cadmium telluride (MCT) detector (~37° normal to substrate). The ice is not subjected to VUV irradiation during the IR measurement period, and this duration is not included in the total irradiation time reported. Following irradiation of the initial ice layer, a second 1.6 μm thick layer of the parent molecule was deposited and irradiated for an additional 3 hours. IR scans are conducted before and after the 3-hour irradiation of this second layer.

**Mm-wave detection.** Following irradiation, the ice was allowed to rest at 20 K for 1 hour before being raised to the BGC which is also held at 20 K by a closed-cycle He cryostat (Sumitomo RDK415D2-F70L). To achieve sufficient sensitivity for detection of molecules subliming from an ice by rotational spectroscopy, CPICE employs buffer gas cooling as previously described.[52] Sublimation is induced by a 50-ohm cartridge heater (Lake Shore Cryotronics, HTR-50) and monitored by a silicon diode temperature sensor (Lake Shore Cryotronics, DT-670C-BO-QT32-2, uncalibrated, ±1 K accuracy) stationed behind the substrate which slowly warms the ice to perform TPD at a ramping rate of 5 K min$^{-1}$. As the molecules reach their sublimation temperature, they enter the gas phase and migrate into the BGC. Within the BGC, the desorbing gas is cooled through elastic collisions with 20 K neon atoms, which flow through the cell at 4 SCCM. At the same time, the CPICE mm-wave spectrometer produces several broadband chirps (25 MHz wide) and free-induction decay (FID) collection cycles to measure the rotational emissions of individual products within the multicomponent gas desorbed from the ice surface. The frequency of the excitation chirp is chosen to target a known high-intensity transition of an expected product based upon previous literature and the observed FTIR abundances. All products identified and quantified by RAIRS and mm-wave spectroscopy for both systems are shown in Tables S1 and S2 of the Supplementary Information, respectively.

**FTIR and mm-wave Quantification** The ratios of the column density of irradiation products in both the gas and ice are determined. The column density of each observed product is calculated in the ice and gas-phase using the IR and mm-wave spectra, respectively. The ice-phase column densities (N) are derived from the IR spectra by dividing the integrated intensity of a vibrational band ($\int \tau(\nu)d\nu$) by the known band strength (A) associated with that mode. As shown in eq. 1.

$$N\ (molecule\ cm^{-2}) = \frac{\int \tau(\nu)d\nu}{A\ (cm\ molecule^{-1})} \qquad \text{Eq. 1}$$

All vibrational modes and band strengths with associated references used for quantification are in Table S1 of the supplementary information. It is known that the reported vibrational band strengths vary significantly within the literature. Therefore, the most recent and applicable values are used for quantification here. A correction factor is calculated by deriving the band strength of the CN st. of the parent molecule using the known column density of the ice and the measured integrated intensity of this transition prior to irradiation. The average band strength of the CN st. of $CH_3CN$ measured from the RAIRS spectrum was 2.5 x 10$^{-18}$ cm molecule$^{-1}$, compared to the literature value of 2.2 x 10$^{-18}$ cm molecule$^{-1}$.[32, 56] A scale factor is applied to correct the band strength to match the literature value. This factor is then applied to the integrated peak intensities of the molecules of interest. Gas-phase column density ratios were calculated using a technique described previously.[51]

## Results

We first consider products observed by RAIRS in the ice, then turn to the products detected in the gas phase following TPD. In the following we chiefly focus on products detected both in the gas phase and in the ice and highlight capabilities of CPICE for isomer-specific detection of sublimed molecules as well the complementary advantages of FTIR and mm-wave identification and quantification of products. Additional radiation-dependent features are seen in the IR spectra but these will be reserved for future investigation.

### Condensed Phase

The parent molecules MeCN and EtCN and their products from VUV photolysis are identified using a combination of IR peaks that appear following irradiation and vanish during TPD as shown in Fig. 2. Due to the sensitivity of vibrational mode frequencies to various physical and chemical properties of the ice, several checks must be conducted to enhance the confidence in band assignment. We will refer to a 5-point check list following Oberg and coworkers.[57]

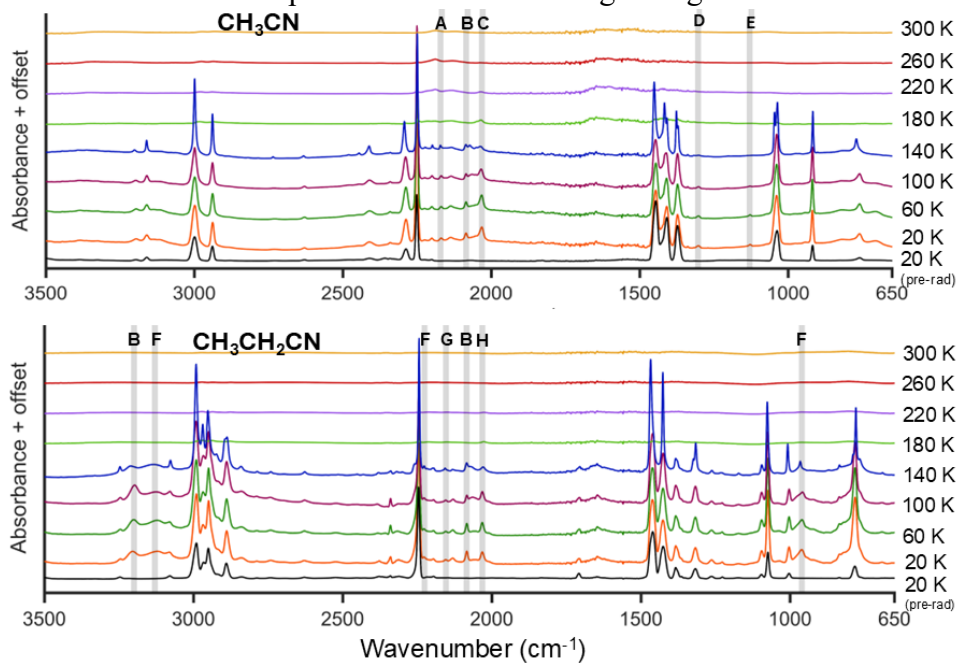

**Figure 2**. FTIR spectra during TPD of irradiated methyl cyanide (top) and ethyl cyanide ice (bottom). The peaks used for product identification are highlighted by grey bars. A: $CH_3NC$ (NC st., 2171 $cm^{-1}$), B: HCN (CH st., 3200 $cm^{-1}$; CN st., 2085 $cm^{-1}$), C: CH2NH (CCN st., 2032 $cm^{-1}$), D: $CH_4$ (CH def., 1302 $cm^{-1}$), E: $CH_3NH_2$ ($CH_3$ rock., 1130 $cm^{-1}$), F: $CH_2CHCN$ (CH st., 3123 $cm^{-1}$; CN st., 2228 $cm^{-1}$; $CH_2C$ + CCHCN wag., 962 $cm^{-1}$), G: $CH_3CH_2NC$ (NC st., 2155 $cm^{-1}$), H: $(CH_3)HCCNH$ (CCN st., 2030 $cm^{-1}$).

1. Observed band position after photolysis are within 15 $cm^{-1}$ of a measured pure ice band, or within 50 $cm^{-1}$ of a calculated band position.

2. The expected band decreases in intensity during the temperature-programmed desorption experiments. This decrease corresponds to the predicted desorption temperature of the species to which the band is assigned.
3. The disappearance of bands corresponding to photolysis fragments reacting as temperature increases.
4. The growth of bands resulting from the recombination of fragments forming other species as temperature increases.
5. Bands shift corresponding to the crystallization of the bulk ice expected from literature at the appropriate temperatures.

The main MeCN and EtCN parent peaks seen in the IR spectra are summarized in Table S1. The products of MeCN photon and ion irradiation have been previously identified using FTIR and are used to assist in peak identification here.[32, 35, 37] Six total isolated vibrational modes of products were observed in the ice-phase following VUV irradiation of MeCN. Specifically, the methylamine $CH_3$ rock (1128 cm$^{-1}$), methane CH def. (1303 cm$^{-1}$), ketenimine CCN st. (2030 cm$^{-1}$), HCN CN st. (2085 cm$^{-1}$), and $CH_3NC$ NC st. (2170 cm$^{-1}$). [32, 33, 37, 58] A summary of observed vibrational bands and associated band strengths used for quantification of these product species are also provided in Table S1 of the supplemental information. The vinyl cyanide CN st. was not seen in the MeCN irradiated ice. EtCN could not be quantified in the MeCN ice owing to interference from the parent molecule, but it is readily observed in the gas phase following TPD. HCN signal is robust and detected from both parent molecules both in the ice and in the gas phase. For MeCN irradiation, methylamine is the most abundant product followed by methane. Methane is not detectable by rotational spectroscopy and methylamine has only weak transitions in our spectrometer window and as such neither of these are observed in the gas phase following TPD. Ketenimine, an isomer of $CH_3CN$, is seen as a minor component in the ice but is not detected in the gas phase, again owing to weak transitions in our spectrometer window.

The column density of all photoproducts for both systems over VUV fluence are shown in Figure 3. It is notable that certain products follow different formation rates during energetic processing.

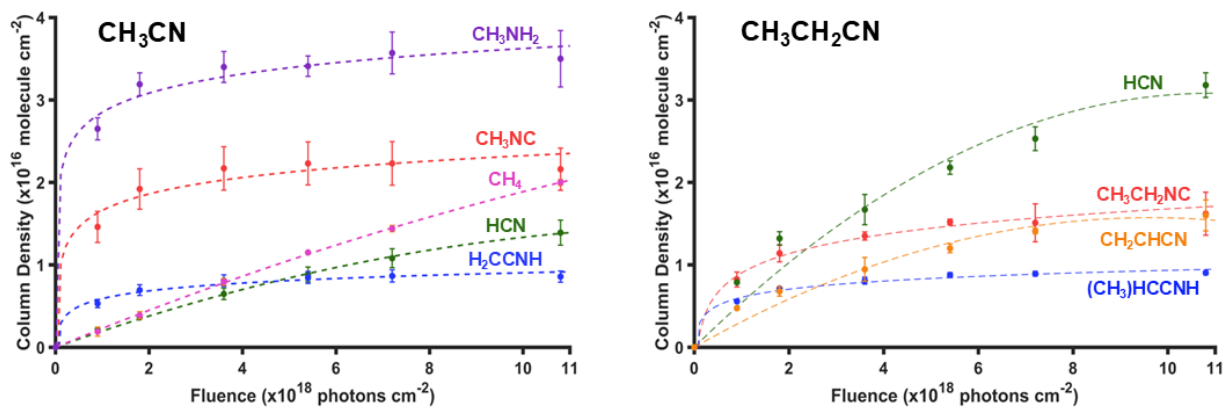

**Figure 3.** Condensed-phase column density of photoproducts of VUV photolysis over time. Dashed lines are approximate trendlines to help guide the reader.

Within the $CH_3CN$ ice, the larger photoproducts observed including $CH_3NH_2$, $CH_3NC$, and $H_2CCNH$ display a logarithmic growth as a function of photon fluence and ultimately reach chemical equilibrium within the irradiation time explored. The smaller photoproducts observed including HCN and $CH_4$ follow relatively slow accumulation within the ice and continue to grow, suggesting their formation and destruction are governed by kinetic limitations or ongoing reaction pathways beyond the irradiation timeframe. The inferred relative abundances of products are given in Table 1 below along with analogous results for EtCN and for the gas phase results following TPD.

|  | Gas Phase | | 20 K Condensed Phase | |
| --- | --- | --- | --- | --- |
| **Species** | **$CH_3CN$** | **$CH_3CH_2CN$** | **$CH_3CN$** | **$CH_3CH_2CN$** |
| **HCN** | 1.5 (0.1) | 5.7 (0.4) | 0.5 (0.1) | 1.7 (0.2) |
| **Ketenimine, Methyl ketenimine** | nd | nd | 0.3 (0.1) | 0.6 (0.1) |
| **Isonitrile (MeNC, EtNC)** | 1 | 1 | 1 | 1 |
| **Methylamine** | nd | nd | 2.3 (0.4) | nd |
| **Methane** | ns | ns | 0.9 (0.2) | nd |
| **Vinyl cyanide** | nd | 2.5 (0.5) | nd | 0.9 (0.1) |
| **Ethyl cyanide** | 1.4 (0.1) | nd | overlap | parent |

**Table 1**. Table of branching ratios from VUV irradiated MeCN and EtCN ices in the condensed phase and following thermal desorption to the gas phase. All molecules are scaled to the isonitrile species. (key: nd: non-detection, ns: not spectroscopically active, overlap: has feature overlapping with parent)

For VUV-irradiated EtCN ice, the major difference is detection of $CH_2CHCN$ (VCN) both in the ice and in the gas phase. In the EtCN case, the analogous methylketenimine is detected in the ice though less abundant than any other product measured. Methane is not detected. HCN is found to be the major product from EtCN VUV photolysis, followed by its isonitrile. As shown in Figure 3, HCN again exhibits a non-logarithmic formation trend across the explored fluence range. In contrast, the isonitrile and imine isomer formed rapidly reach equilibrium, following a logarithmic behavior similar to that observed for the isomerization products of MeCN. VCN displays a formation pattern comparable to that of HCN.

### Gas phase

TPD spectra with mm-wave detection for the processed MeCN and EtCN ices are shown in Fig. 4. Three photoproducts were observed and quantified in the gas phase following TPD of VUV irradiated MeCN ices, all of which are nitriles. This includes HCN (J = 1 – 0 centered at 88.6316 GHz), $CH_3NC$ (J = J = $4_1 - 3_1$ at 80.4200 GHz and J = $4_0 - 3_0$ at 80.4219 GHz), and EtCN (J = 9 $0,9 - 8_{0,8}$ at 79.6775 GHz and J = $10_{0,10} - 9_{0,9}$ at 88.3237 GHz). Similar to MeCN, three nitriles were also observed in the sublimation of irradiated EtCN ices. HCN was probed using the same J = 1-0 transition mentioned previously. VCN (J = $9_{1,8} - 8_{1,7}$ at 87.3128 GHz) and $CH_3CH_2NC$ (J = $9_{2,7} - 8_{2,6}$ at 88.0743 GHz) were also observed. A summary of all rotational transitions, 18 K intensities, and line strength factors are provided in Table S2 of the supplementary information.

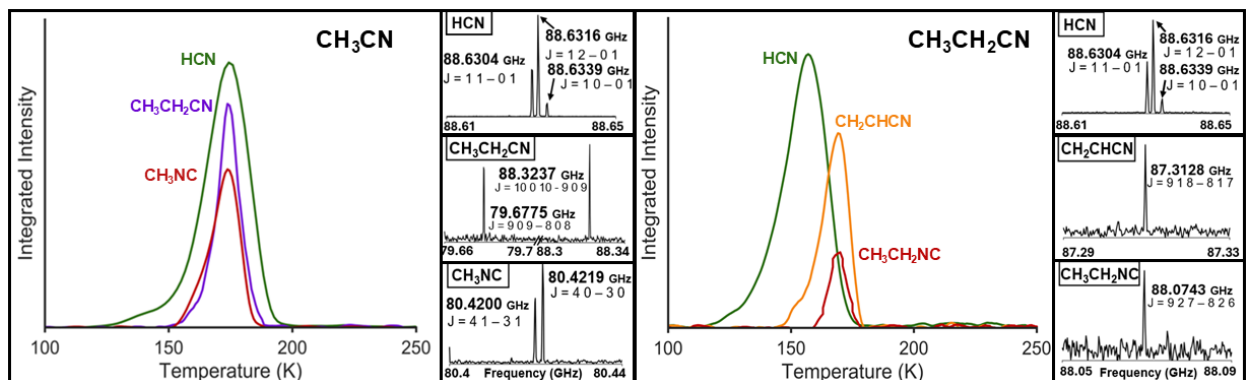

**Figure 4.** TPD spectra of the frequency integrated signal for photoproducts observed in the gas-phase following sublimation of VUV irradiated cyanide ices. Inset: mm-wave spectra of species during peak desorption temperature. TPD curves are corrected to reflect product branching ratios and mm-wave spectra are corrected for power and line strength factors.

Thus, in all cases we detected the isonitrile of the parent molecule, both in the ice and in the gas phase. As these likely arise from direct isomerization of the parent molecule in analogous processes for both species (see below), we find it useful to normalize all spectra to these abundances. For MeCN, the HCN : $CH_3NC$ ratio is 1.5:1 in the gas phase, but 0.5:1 in the ice. For EtCN irradiation, the HCN: $CH_3CH_2NC$ ratio is considerably enhanced relative to the MeCN case: 5.7:1 in the gas phase and 1.7:1 in the ice. We consider the gas phase quantification more accurate, as RAIRS can be complicated by interference effects that are not accounted for here.[59] It is interesting that in both cases the relative abundance of HCN is lower by a factor of 3 in the ice compared to the gas phase. The other important product seen both in the gas phase and the ice is VCN from EtCN irradiation. Here the gas phase VCN: $CH_3CH_2NC$ ratio is 2.4 and in the ice 0.9, a factor of 2.7. As we have normalized everything to the isonitriles, we could achieve good agreement between all gas phase and ice phase relative abundances if we increase the assumed band strength of both isonitriles by a factor of ~3. This is discussed further below.

**Discussion**

We now consider some of the implications of these observations for the underlying chemistry. Within our photon energy range, the photo-isomerization of MeCN can occur via the following pathways:

$CH_3CN + hv \rightarrow CH_3NC$ (5)
$CH_3CN + hv \rightarrow CH_2CNH$ (6).

Unimolecular isomerization of $CH_3NC$ to MeCN has been studied extensively both by experiment and high-level electronic structure methods.[60, 61] The results suggest a TS energy of 2.72 eV for the reverse reaction which proceeds by 1,2 methyl migration. This is far below the $CH_3$ + CN dissociation threshold of 5.24 eV.[62] Shock tube studies of the isomerization of ketenimine to MeCN give an activation energy of 3.04 eV.[63] Combined with the relative energy of ketenimine (1.18 eV at the G4 level)[64] this suggests the isomerization of MeCN to ketenimine in this case is very near the 4.12 eV homolytic bond dissociation threshold[62], consistent with a roaming-like mechanism as suggested by Wentrup.[63, 65] These are gas phase numbers but likely serve as an upper limit for the energetics in the ice. It may be seen clearly in this work and other laboratory studies mentioned above that these isomers readily form when neat MeCN ices are exposed to VUV irradiation. In a matrix isolation study coupled with DFT calculations, Cho investigated the formation of $CH_3NC$, $CH_2CNH$ and $CH_2NCH$ following laser-ablation of MeCN at 10 K.[66] Hudson and Moore showed both isonitrile and ketenimine formation with the latter dominating for proton-irradiated ices and the former for VUV, consistent with our findings.[32] Isonitrile formation was quenched in the presence of water ice; instead $OCN^-$ was found. Both isomerization transition state energies are well below the photon energies produced by our lamp. Mencos and Krim find an isonitrile to imine ratio of 10.4 once again finding a larger abundance of the isonitrile form when irradiating MeCN ices with energetic N ions.[67] We find the MeNC:ketenimine ratio of about 1:0.3 following VUV irradiation of our ices. In our measurements, both products are seen to have reached chemical equilibrium within our VUV radiation time. This difference seen in both EtCN and MeCN isomerization from our work compared to other work at lower temperatures requires further systematic investigation of its dependence on wavelength and dose.

The product branching ratio of gas phase MeCN photolysis at wavelengths below 235 nm show two primary pathways:[29, 68]

$CH_3$ + CN (20 %)
$CH_2CN$ + H (80 %)

At low temperatures in the ice, H atoms have greater mobility than the other fragments formed, leading to an expected significant contribution to chemistry within ices. In addition, H atoms formed by molecular dissociation reactions have high average translation energy that can overcome hydrogenation reaction barriers. Assuming the resulting radical inherits all the energy remaining following photon absorption and bond breaking, an H atom produced from fragmentation of one of the C-H bonds of MeCN has kinetic energies up to 3 eV (7 eV – 4 eV).[53, 68, 69] This has implications for the formation of several of the products observed as discussed in the following.

Methylamine ($CH_3NH_2$) is the most abundant species we see in the irradiation of MeCN in the condensed phase. Although not reported by Hudson and Moore, it is also seen as the dominant FTIR-detected product in X-ray irradiated ices.[35] It has been observed in multiple interstellar molecular clouds.[70, 71] Notably, however, it is not seen in Titan's atmosphere.[30] As an important intermediate for the formation of glycine on ice surfaces[72] and its relevance for astrobiological evolution, there are many proposed formation pathways for methylamine in both gas and surface grain reactions. The dominant reaction proposed by Garrod is the radical reaction between $CH_3$ and $NH_2$.[72] While the $CH_3$ radical is expected to be very abundant in VUV irradiated MeCN, the production of $NH_2$ is not a favorable pathway following the photolysis of MeCN under the conditions investigated in this experiment as there is no evidence of the 1200 cm$^{-1}$ vibrational mode of $NH_2$ at 20 K following irradiation. Theule et al. suggested the successive hydrogenation of HCN that leads to the formation of $CH_3NH_2$.[73] This hydrogenation process is a favorable pathway within VUV irradiated MeCN ices due to the large abundance of both energetic H atoms and the presence of HCN.

The mobile H atoms also allow for a high probability of H + $CH_3$ interaction to form methane. Methane can also form from radical chain reactions first by C-C bond fission in MeCN, then by $CH_3$ + $CH_3CN$ giving $CH_4$ + $CH_2CN$.[74] This yields the CN radical, methane, and a newly formed $CH_2CN$ radical. The $CH_2CN$ radical is an important intermediate for the formation of EtCN that we see subliming from our processed MeCN ice. EtCN can form from the reaction of $CH_2CN$ with a nearby $CH_3$ radical. The probability of these two species interacting increases with temperature due to increased radical mobility as the radical chain reactions propagate through the ice. These likely serve as the precursor to VCN formation where hot EtCN formed by $CH_3$+$CH_2CN$ can eliminate $H_2$ to form VCN.

The pure EtCN ice FTIR spectra contained several peaks associated with the fundamental vibrational modes of the parent as given in Table S1 of the supplementary information. VCN is only unambiguously observed in the RAIRS spectrum of VUV irradiated EtCN by its $v_{CN}$ stretch at 2228 cm$^{-1}$ which is not present in the irradiated MeCN spectrum. While the C-H st. of VCN could contribute to the broad IR feature found between 3200-3100 cm$^{-1}$, the C-H stretches of other product species are expected to overlap in this region as well.[75] Our own calculations for the isolated molecules suggest a factor of 12 larger cross section for isoniriles than the nitrile on the C-N stretch band. This may be reduced or broadened in the ice to give the factor of three we infer. This revision could bring all of the gas and condensed phase values observed in this work into agreement.

The VUV photochemistry of EtCN is not known, but its gas phase pyrolysis has been studied and provides useful insight. In the absence of radical scavengers, EtCN pyrolysis gives VCN + $H_2$ and HCN + $C_2H_4$ with activation energies of 2.56 eV and 2.94 eV, respectively.[76] Formation of these products is inhibited in the presence of radical scavengers, suggesting underlying these observations is the primary importance of radical chain reactions initiated by C-C bond fission giving $CH_3$ + $CH_2CN$. Successive H abstraction of a neighboring EtCN molecule by $CH_3$ or $CH_2CN$ then gives VCN, although we do not see methane formation in the EtCN case. HCN formation follows from H + $CH_2CN$ → HCN + $CH_2$ [77], which explains why it is enhanced for EtCN.

In EtCN ices, we observe a slow growth of VCN over the irradiation time of the experiment. The abundance of this product in the MeCN ice is dependent upon the abundance of EtCN. If the latter forms more readily at warmer temperatures during TPD (when there is no VUV irradiation), VCN production will be limited. This and the ease with which VCN forms photopolymers at this wavelength region can explain why it is not observed in the mm-wave TPD experiment for MeCN.

EtCN can isomerize into either the isonitrile form ($CH_3CH_2NC$) or into an imine form (($CH_3$)HCCNH) under energetic processing.[32, 75]. Both forms are observed here. Branching ratios of 8%, 5%, 2% for $CH_3CH_2NC$, $CH_2CHCN$, and methylketenimine have been reported for EtCN by Couturier-Tamburelli et al. at when irradiating their ice at 95 K. This closely aligns with the relative abundances we observe under our conditions, despite differences in lamp wavelength distribution, flux, irradiation time, and ice temperature. Our total dose for EtCN irradiation is 195 eV/molecule, which is within the same order of magnitude as Couturier-Tamburelli (144 eV/molecule). However, we observe only ~ 9% depletion of our EtCN, while they observe 54% depletion over the course of 420 mins at a flux of ~$5 \times 10^{14}$ photons $cm^{-2}$ $s^{-1}$ at 20 K.

Under the same irradiation conditions, we observe a greater fraction of EtCN depletion compared to MeCN, consistent with a larger VUV absorption cross section for EtCN. Depletion is calculated by the difference in the integrated intensities of the parent $v_{CN}$ stretch peak before and after irradiation. The exact absorption cross section for both species in the condensed phase has not been quantified, but the relative absorbance between both species shows a greater VUV absorption cross section in the most intense 160 nm wavelength region of our lamp.[78] It has also been reported that the gas phase low energy electron ionization cross section is much greater in EtCN than MeCN. However, we do not expect direct ionization from the $D_2$ lamp as we are irradiating at energies well below the ionization potentials of EtCN (11.84 eV) and MeCN (12.20 eV).[79] Nevertheless, charge transfer reactions induced by VUV irradiation below ionization potentials of these species in the condensed phase could lead to ionization and photochemistry.[80, 81]

The increased abundance of CN radicals in MeCN ice could cause an increased rate of refractive material formation. This was observed as a brown residue that was stable at room temperature following VUV irradiation of MeCN. There was little to no residue observed following EtCN irradiation, which suggests that the polymerization loss channel of CN is less prominent in the longer chain nitrile ice.

While we detect methylamine in the condensed phase, we do not observe it in the gas phase following TPD. It has a unique desorption temperature relative to the other photoproducts observed in the $CH_3CN$ thermal desorption experiments. In pure $CH_3NH_2$, the desorption temperature is expected to be ~ 110 K.[82] The IR signal corresponding to $CH_3NH_2$ disappears between 100 – 110 K within our RAIRS TPD spectra. This lack of detection following sublimation can be attributed to the combination of uncertainties in band strength due to the influence of the local ice environment of the measured molecule, and the rotational line intensity of the available rotational transitions in our frequency region that are nearly a factor of 70 weaker than that of any other product that was observed.[83, 84]

## Conclusion

This work demonstrates that the combination of RAIRS and broadband mm-wave detection of molecules provides a powerful approach to quantitatively link the chemistry occurring in VUV-irradiated ices to the gas-phase products observed after TPD. Through examination of MeCN and EtCN ices in situ, several photoproducts including HCN, isocyanides, ketenimines, methylamine, methane, and VCN were observed. The relative abundances of gas-phase and ice-phase products were found to be consistent when the IR band strength of the isocyanide C-N stretch was assumed to be ~ 3 times greater than that of the cyanide. These validation experiments open the door for future work targeting more complex or less characterized ice systems in an effort to expand our understanding of molecular evolution in cold astrophysical environments. This integrated approach will be applied to explore isotope effects, the role of co-deposited species, and reaction kinetics, offering a comprehensive framework for connecting solid-phase chemistry with observable gas phase signatures in space.

## Acknowledgements

We gratefully acknowledge the National Science Foundation Advanced Manufacturing Program, Grant No. 2314347, and The Cottrell Scholar Award No. CS-CSA-2024-016 from the Research Corporation for Science Advancement. We thank Arthur G. Suits for many helpful discussions.

## Supporting Information
Table S1 with vibrational modes, frequencies, and band strengths. Table S2 with rotational transitions and frequencies, lines strength factors, 18 K intensities.